\documentclass[12pt]{article}
\usepackage{amssymb}
\usepackage{amsmath}
\usepackage{latexsym}

\date{\nonumber}
\newtheorem{prop}{Proposition}
\newtheorem{lemma}{Lemma}

\def \be  {\begin{equation}}
\def \ee  {\end{equation}}
\def \part {\partial }

\begin{document}
\setlength{\baselineskip}{16pt}

\renewcommand{\theequation}{\arabic{section}.\arabic{equation}}
\renewcommand{\thesection}{\arabic{section}.}

\

\section*{\Large\bf 
Higher order potential expansion 
for the continuous limits of the Toda hierarchy
}

\

\parbox[b]{1.2truecm}{\quad}
\parbox[b]{13truecm}{
Runliang Lin\dag\S\footnotemark, 
Wen-Xiu Ma\ddag\footnotemark\  and  
Yunbo Zeng\dag\footnotemark

{\small\dag {\ Department of Mathematical Sciences, Tsinghua University,
Beijing 100084, P.R. China} }\\
{\small\S {\ Service de physique de l'\'etat condens\'e, CEA-Saclay,
F-91191 Gif-sur-Yvette Cedex, France} }\\
{\small\ddag {\ Department of Mathematics, City University of Hong Kong,
Hong Kong, P.R. China} }\\ 
{\small \ddag {\ Department of Mathematics, University of South Florida, 
Tampa, FL 33620-5700, USA} }
}
\renewcommand{\footnoterule}{\rule{10pt}{0pt}\vspace{0pt}}
\addtocounter{footnote}{-2}\footnotetext{E-mail address: rlin@math.tsinghua.edu.cn}
\addtocounter{footnote}{1}\footnotetext{E-mail address: mawx@math.cityu.edu.hk}
\addtocounter{footnote}{1}\footnotetext{E-mail address: yzeng@math.tsinghua.edu.cn}

\vspace{0.5truecm}

\parbox[b]{1.2truecm}{\quad}
\parbox[b]{13truecm}{\small
{\bf Abstract. } 
A method for introducing the higher order terms in the potential expansion 
to study the continuous limits of the Toda hierarchy
is proposed in this paper.
The method ensures that the higher order terms are differential polynomials of the lower ones
and can be continued to be performed indefinitly.
By introducing the higher order terms, the fewer equations in the Toda hierarchy 
are needed  in the so-called recombination method to recover the KdV hierarchy.
It is shown that the Lax pairs,
the Poisson tensors, and the Hamiltonians of the Toda hierarchy
tend towards the corresponding ones of the KdV hierarchy in continuous limit.
}

\vspace{0.5truecm}

\section{Introduction}

The continuous limits of discrete systems 
are one of the remarkably important research areas
in soliton theory 
\cite{Toda-73,Case-74,Kupershmidt-85,Ablowitz-81}.
In recent years, more attention was focused on
the continuous limit relations 
between hierarchies of discrete systems and
hierarchies of soliton equations 
\cite{Zeng-95a,Zeng-95b,Morosi-96,Morosi-98a,Morosi-98b}.
The so-called recombination method, i.e., 
properly combining the objects (such as the vector fields) of discrete systems, 
was first proposed to study the continuous limit of 
the Ablowitz-Ladik hierarchy \cite{Zeng-95a} and
the Kac-van Moerbeke hierarchy \cite{Zeng-95b}.
Morosi and Pizzocchero also used the recombination method
to study the continuous limits of some integrable lattices 
in their recent works  
\cite{Morosi-96,Morosi-98a,Morosi-98b}. 
Up to now, there has not been much work concerning the continuous limit relations
between lattices and differential equations,
which have different numbers of potentials.
Furthermore, to the best of our knowledge, 
there is no work which successfully
gives a way to introduce the higher order terms in potential expansion 
to study the continuous limit relations 
between hierarchies of lattices and 
hierarchies of soliton equations.
Illumined by Gieseker's conjecture \cite{Gieseker},
we will propose a method for finding the higher order terms in potential expansion
to study the continuous limit relation between the Toda hierarchy and the KdV hierarchy
by the recombination method.

In 1996, Gieseker proposed a conjecture \cite{Gieseker}:

{\parindent=0pt\bf Conjecture.}  
{\it Denote $w(n,t)$ and $v(n,t)$, where $n\in\mathbb Z$ and $t\in\mathbb R$,
are the two potentials of the Toda hierarchy,
and let $f$ be a function of $x\in \mathbb R$ and $t\in \mathbb R$.
There are $\Phi_i(f)$'s, which are the differential polynomials of $f$,
so that if we define
\begin{subequations}
\label{w-v-q}
\begin{equation}
	w(n,t)=-2+ f(x,t) h^2+h^2\sum_{i=1}^L \Phi_i(f(x,t)) h^i,
\end{equation}
\begin{equation}
	v(n,t)=1+ f(x,t)h^2-h^2\sum_{i=1}^L \Phi_i(f(x,t)) h^i,
\end{equation}
\end{subequations}
where $h$ is the small step of lattice and $x=nh$,
then by taking suitable linear combinations of the equations of Toda hierachy
under the definition (\ref{w-v-q}),
we can produce asymptotic series whose leading terms in $h$ are the KdV hierarchy 
if $L$ is large enough. }

In \cite{Gieseker}, 
Gieseker proposed a way to introduce $\Phi_i(f)$ by using the Toda lattice
\begin{equation}
\label{toda}
	w_t= v-Ev=v-v^{(1)}, \qquad v_t=v(E^{(-1)}w-w)=v(w^{(-1)}-w),
\end{equation}
where the shift operator $E$ is defined by
        $$ (E f) (n) = f(n+1), \quad 
        f^{(k)} (n) =E^{(k)} f (n)=f(n+k), \quad n, k\in {\mathbb Z}. $$
For instance, in order to find $\Phi_1(f)$, 
substituting the definition (\ref{w-v-q}) into the equation (\ref{toda}) 
and expanding the shift terms out by Taylor's theorem
\begin{subequations}
\begin{equation}
\label{Phi_1a}
	\frac{d f}{dt} + \frac{d \Phi_1(f)}{dt} h = 
	-\frac{d f}{dx} h -  \frac{d^2 f}{2dx^2} h^2 + \frac{d \Phi_1(f)}{dx} h^2 +O(h^3),
\end{equation}
\begin{equation}
	\frac{d f}{dt} - \frac{d \Phi_1(f)}{dt} h = 
	-\frac{d f}{dx} h + \frac{d^2 f}{2dx^2} h^2 - \frac{d \Phi_1(f)}{dx} h^2 +O(h^3).
\end{equation}
\end{subequations}
Combining the above two equations we know
\begin{equation}
	\frac{d f}{dt} = -\frac{d f}{dx} h + O(h^3),
\end{equation}
then by the chain rule we have
\begin{equation}
	 \frac{d \Phi_1(f)}{dt}  = -\frac{d \Phi_1(f)}{dx} h + O(h^2),
\end{equation}
Notice the above equation and the equation (\ref{Phi_1a}) one can get
\begin{equation}
	\frac{d \Phi_1(f)}{dx} = \frac 1 4 \frac{d^2 f}{dx^2},
\end{equation}
by integration it yields
\begin{equation}
	\Phi_1(f) = \frac 1 4  \frac{d f}{dx}.
\end{equation}
We can see that the integration must be used in this process for finding $\Phi_i(f)$.
As a consequence, there is a problem that whether this process can be continued indefinitely
and the $\Phi_i(f)$'s, found in this process, are the differential polynomials of $f$. 

The Gieseker's conjecture were proved 
in the following three cases of (\ref{w-v-q}) \cite{ZLC}:  
	$$ \mbox{(a)} \qquad L=0, \quad f(x,t)=\frac12 q(x,t); $$
	$$ \mbox{(b)} \qquad L=1, \quad f(x,t)=\frac12 q(x,t), \quad \Phi_1(f)=\frac{1}{8} q_x; 
$$
	$$ \mbox{(c)} \qquad L=2, \quad f(x,t)=\frac12 q(x,t), \quad \Phi_1(f)=\frac{1}{8} q_x, 
		\quad \Phi_2(q)=-\frac{1}{32} q^2. $$
It was found that the fewer equations in the Toda hierarchy are needed in the recombination 
method
for the case (c)  to give the KdV hierarchy than for the case (a).

In this paper, we will give a new method to 
introduce $\Phi_i(f)$ required in (\ref{w-v-q}) instead of the Gieseker's process 
in order that we can derive the continuous limit relation 
between the Toda hierarchy and the KdV hierarchy
by the recombination method.
Following our approach for finding $\Phi_i(f)$, 
one can easily see that 
the $\Phi_i(f)$'s are all differential polynomials of $f$.
Compared with the previous work in  \cite{ZLC},
we will show that the fewer equations in the Toda hierarchy are needed 
in the recombination method
for giving the KdV hierarchy
if higher order terms are introduced in the potential expansion (\ref{w-v-q}).
We will also present that the Lax pairs, the Poisson tensors, 
and the Hamiltonians of the Toda hierarchy
tend towards the corresponding ones of the KdV hierarchy in continuous limit.

\section{Basic notation and some known results}

For latter use, we list some notation and results in \cite{ZLC}.
Let us consider the following
discrete isospectral problem \cite{TUTD,Newell},
\begin{equation}
\label{TDEIGEN}
        L y = (E + w + v E^{(-1)}) y =\lambda y,
\end{equation}
where $w=w(n,t)$ and $v=v(n,t)$ depend on integer $n\in {\mathbb Z}$
and real variable $t \in {\mathbb R},$ and $\lambda$ is the spectral parameter.

The equation in the Toda hierarchy associated with (\ref{TDEIGEN}) 
can be written as
the following Hamiltonian equation \cite{TUTD}
\begin{equation}
\label{TDHIERARCHY}
        {\left( \begin{array}{c} w \\ v \end{array} \right)}_{t_m}
        =J K_{m+1}=J \frac{\delta H_{m+1}}{\delta u}, \quad 
	m=0,1,...,
\end{equation}
where $\frac{\delta}{\delta u}=
	{(\frac{\delta}{\delta w}, \frac{\delta}{\delta v})}^{T},$
and the Poisson tensor $J$ and the Hamiltonians $H_i$ are defined by
        $$ J\equiv \left( \begin{array}{cc}
                0 & J_{12} \\ J_{21} & 0 \end{array} \right)
        \equiv \left( \begin{array}{cc}
                0 & (1-E) v \\ v(E^{(-1)} -1) & 0 \end{array} \right),$$
\begin{equation}
\label{Ki}
        K_i\equiv \left(\begin{array}{c} K_{i,1} \\ K_{i,2} \end{array} \right)
	=\frac{\delta H_{i}}{\delta u}=
        \left( \begin{array}{c} -b_i^{(1)} \\ \frac{a_i}{v} \end{array}
                \right), \quad i=0,1,..., 
\end{equation}
        $$ H_0=\frac{1}{2}\ln v, \quad H_i=-\frac{b_{i+1}}{i},  \quad      
                i=1,2,..., $$
with $a_0=\frac{1}{2},$ $b_0=0,$ and
\begin{equation}
\label{ab}
        b_{i+1}^{(1)}=w b_i^{(1)} -(a_i^{(1)}+a_i), \qquad
        a_{i+1}^{(1)}-a_{i+1} = w(a_i^{(1)}-a_i)+vb_i-v^{(1)}b_i^{(2)},
\end{equation}
for $i=0,1,....$
The Lax pairs for the $m$th equation of the Toda hierarchy (\ref{TDHIERARCHY})
are given by (\ref{TDEIGEN}) and 
\begin{equation}
	y_{t_m} = A_m y = \sum_{i=0}^m (- v b_i^{(1)} E^{(-1)}- a_i )
	(E + w + v E^{(-1)})^{m-i} y, \qquad m = 0,1,....
\end{equation}
The equations (\ref{TDHIERARCHY}) have the bi-Hamiltonian formulation
\begin{equation}
\label{TDBIHS}
        G K_{i-1} = J K_i, \quad i=1,2,...,
\end{equation}
        $$ G \equiv \left( \begin{array}{cc}
        vE^{(-1)}-v^{(1)}E & w(1-E)v \\
        v(E^{(-1)}-1)w  &  v(E^{(-1)}-E)v \end{array} \right), $$
where $G$ is the second Poisson tensor.
The Toda hierarchy also has a tri-Hamiltonian formulation and
a Virasoro algebra of master symmetries \cite{Ma-JMP,Ma-99}. 
The first four covariants $K_i$'s are
\begin{equation}
        K_0 = \left( \begin{array}{c}
        0 \\ \frac{1}{2v} \end{array} \right), \quad
        K_1 = \left( \begin{array}{c}
        1 \\ 0 \end{array} \right), \quad
        K_2 = \left( \begin{array}{c}
        w \\ 1 \end{array} \right), \quad
        K_3 = \left( \begin{array}{c}
        v+v^{(1)}+w^2 \\ w+w^{(-1)} \end{array} \right).
\end{equation}

The Schr\"{o}dinger spectral problem is given by
\begin{equation}
\label{KDVEIGEN}
	\overline L \overline y = (\partial^2_x+q) \overline y 
	= -\overline\lambda \overline y.
\end{equation}
which is associated with the KdV hierarchy \cite{Newell}
\begin{equation}
\label{KDVHIERARCHY}
        q_{t_m} = B_0 P_m=B_0 \frac{\delta \overline H_m}{\delta q}, 
        \qquad m=0,1,...,
\end{equation}
where the vector field possesses the bi-Hamiltonian formulation with two
Poisson tensors $B_0$ and $B_1$
\begin{equation}
\label{KdVBIHS}
        B_0  P_{k+1} = B_1 P_k, \quad k=0,1,...,
\end{equation}
        $$  B_0 = \partial\equiv\partial_x, \quad
        B_1 = \frac{1}{4} \partial^3+q \partial+
                \frac{1}{2} q_x, \quad 
        \overline H_i = \frac{4 \bar b_{i+2}}{2i+1},\quad 
        i = 0,1,...,$$
with $\bar b_0=0,$ $\bar b_1=1,$ and
        $$\bar b_{i+1}=(\frac{1}{4}\partial^2+q-
        \frac{1}{2}\partial^{-1}q_x)\bar b_i,
        \quad i=0,1,..., $$
where $\partial^{-1}\partial=\partial\partial^{-1}=1$. 
The first three covariants $P_k$'s read as
\begin{equation}
        P_0 = 2, \quad P_1=q, \quad
        P_2 = \frac{1}{4}(3{q}^2+q_{xx}).
\end{equation}
The well-known KdV equation is the second one:
\begin{equation}
\label{KDV}
        q_{t_2}=\frac{1}{4}{(3 q^2+q_{xx})}_x.
\end{equation}
The Lax pairs for the $m$th equation of the KdV hierarchy (\ref{KDVHIERARCHY})
are given by (\ref{KDVEIGEN}) and
\begin{equation}
\label{KdV-Lax-t}
	\overline y_{t_m} = \overline A_m \overline y
	= \sum_{i=0}^m (- \frac{1}{2}\overline b_{i,x} + \overline b_i \partial)
	(\partial^2+q)^{m-i} \overline y, \qquad m=0,1,....
\end{equation}

Let us consider the Toda hierarchy on a lattice with a small step $h$. 
We interpolate the sequences $(w(n))$ and $(v(n))$ with two smooth 
functions of a continuous variable $x$, 
and relate $w(n)$ and $v(n)$ to $f(x)$ by using (\ref{w-v-q}).
Suppose
	$$ E^{(k)} w(n) = -2+f(x+kh)h^2
		+h^2\sum_{i=1}^L \Phi_i(f(x+kh)) h^i, $$
	$$ E^{(k)} v(n) = 1+f(x+kh)h^2
		-h^2\sum_{i=1}^L \Phi_i(f(x+kh)) h^i,
	\qquad k\in\mathbb Z. $$

In \cite{ZLC}, we got the following result. 

\begin{prop}
\label{prop-spectral}
Under the relation (\ref{w-v-q}) with $f(x,t)=\frac 12 q(x,t)$,
the Lax operator of the Toda hierarchy
goes to the Lax operator of the KdV hierarchy 
in continuous limit, i.e., we have
\begin{equation}
	L = \overline L h^2 + O(h^3),
\end{equation}
\end{prop}

\begin{lemma}
\label{lemma-Ki}
Under the relation (\ref{w-v-q}),
we have
\begin{equation}
\label{KIEXPAND}
	K_i = \left( \begin{array}{c} -b_i^{(1)} \\ \frac{a_i}{v} \end{array}
		\right)
	=\left( \begin{array}{c} \alpha_i \\ \gamma_i \end{array}
	\right) + O(h), \qquad i=0,1,...,
\end{equation}
where $\alpha_i$ and $\gamma_i$ are given by
\begin{subequations}
\label{w1}
\begin{equation}
\label{w1-0-1}
	 \alpha_0 = 0, \qquad \alpha_1 = 1, \qquad 
	   \gamma_0=\frac{1}{2}, \qquad \gamma_1=0, 
\end{equation}
\begin{equation}
	    \alpha_i = (-1)^{(i-1)} C_{2i-2}^{i-1}, 
	  \qquad
		\gamma_i = (-1)^i C_{2i-2}^{i}, 
	\qquad i=2,3,.... 
\end{equation}
\end{subequations}
\end{lemma}

Define 
$\widetilde J= \left( \begin{array}{cc} 0 &\widetilde J_{21} 
                        \\\widetilde J_{12} & 0
                        \end{array} \right)$ 
by requiring that $J\widetilde J=I.$
Then the following lemma is true.

\begin{lemma}
\label{lemma-TKi}
Under the relation (\ref{w-v-q}),
we have
\begin{equation}\label{w2}
        TK_i \equiv\widetilde JG K_i=K_{i+1}+\delta_{i+1} K_0, \quad
        i = 0,1,...,
\end{equation}
where
\begin{equation}
\label{w3}
	 \delta_i = -2(\alpha_i+\gamma_i)
	= (-1)^i \frac{2}{i} C_{2i-2}^{i-1}, \quad i=1,2,.... 
\end{equation}
\end{lemma}

\begin{prop}
\label{prop-Poisson}
Under the relation (\ref{w-v-q}) with $f(x,t)=\frac 12 q(x,t)$,
the Poisson tensors of the Toda hierarchy go to 
those of the KdV hierarchy in continuous limit,
\begin{equation}
\label{JWEXPAND}
        J = - B_0 \left( \begin{array}{cc} 0 & 1 \\ 1 & 0 
                \end{array} \right) h + O(h^2), \quad
        W_{ij}+W_{kl} = -B_1 h^3 +O(h^4), 
\end{equation}
where $W\equiv \frac{1}{4}G\widetilde JG+G=(W_{ij}),$ 
$1\leq i,j \leq 2,$ and
	$$(i,j,k,l)\in 
	\left\{ (1,1,1,2), (1,1,2,1), (1,2,2,2), (2,1,2,2) \right\}.$$
\end{prop}

\section{Higher order potential expansion 
and the continuous limits of the Toda hierarchy}

Now, we give a new method to introduce $\Phi_i(f)$ required in (\ref{w-v-q}) 
and derive the continuous limits of the Toda hierarchy
under the relation (\ref{w-v-q}) with $f(x,t)=\frac 12 q(x,t)$.

\begin{lemma}
\label{lemma-T}
Define the operator as
\begin{equation}
\label{TDEF}
        T\equiv \widetilde JG 
                = \left( \begin{array}{cc} T_{11} & T_{12} \\ 
			T_{21} & T_{22}
                        \end{array} \right).
\end{equation}
Then under the relation (\ref{w-v-q}) with $f(x,t)=\frac 12 q(x,t)$,
the operator $T$ has the following expansions for its entries:
        $$ T_{11} =  -2 + \frac{1}{2} h^2 q + O(h^3), \quad
        T_{12} = 2 + h\partial +(\frac{1}{2}\partial^2+q)h^2+O(h^3),  $$
        $$ T_{21}=2- h\partial +(\frac{1}{2}\partial^2-
                \frac{1}{2}\partial^{-1}q_x)h^2
                +O(h^3), \quad 
        T_{22} = -2+ \frac{1}{2}h^2\partial^{-1}q\partial +O(h^3). $$
\end{lemma}

{\parindent=0pt\it Proof. } The result can be found in \cite{ZLC} 
(see the proof of Lemma 3 in \cite{ZLC}).

\begin{lemma}
\label{lemma-Ki-expand}
Under the relation (\ref{w-v-q}) with $f(x,t)=\frac 12 q(x,t)$,
we have the following expansions,
	$$ K_i\equiv \left(\begin{array}{c}K_{i,1} \\ K_{i,2} \end{array} 
	\right) $$
\begin{equation}
\label{Ki-expand}
	=\left(\begin{array}{c}
	\alpha_{i}+\Psi_{i,1,0}(q)h^2
		+h^2\sum\limits_{j=1}^L 
		h^j(\zeta_{i,1}\Phi_{j}+\Psi_{i,1,j}(q,\Phi_1,...,\Phi_{j-1})) \\
	\gamma_{i}+\Psi_{i,2,0}(q)h^2
		+h^2\sum\limits_{j=1}^L 
		h^j(\zeta_{i,2}\Phi_{j}+\Psi_{i,2,j}(q,\Phi_1,...,\Phi_{j-1})) 
	\end{array} \right)+O(h^{L+3}), 
\end{equation}
for $i=0,1,2,...,$ 
where $\alpha_{i}$ and $\gamma_{i}$ are given in Lemma \ref{lemma-Ki}, 
	$$ \zeta_{0,1}=0,\quad \zeta_{0,2}=\frac{1}{2},\quad
	\zeta_{1,1}=0, \quad \zeta_{1,2}=0,  $$
\begin{equation}
\label{zeta}
	\zeta_{i+1,1}=-2\zeta_{i,1}+2\zeta_{i,2}+\alpha_i-2\gamma_i, \quad
	 \zeta_{i+1,2}=2\zeta_{i,1}-2\zeta_{i,2}+\alpha_i-\frac{1}{2}\delta_{i+1}, 
	\quad i=0,1,..., 
\end{equation}
$\Psi_{i,1,j}(q,\Phi_1,...,\Phi_{j-1})$ stands for the term
which is a differential polynomial of 
$q$, $\Phi_1$, ..., $\Phi_{j-1}$, and etc.
\end{lemma}

{\parindent=0pt\it Proof.} Define $c_i=-vb_i^{(1)},$ $i=0,1,....$
Using the identity \cite{TUTD}
	$$ \sum_{i=0}^k(a_ia_{k-i}+b_ic_{k-i})=0, \qquad k=1,2,..., $$
we can show by the mathematical induction that $a_i$, $b_i$, $c_i$, $i=0,1,...,$
are polynomials of 
$w$, $v$, $w^{(-1)}$, $v^{(-1)}$, $w^{(1)}$, $v^{(1)}$, $...$.
According to the definition of $K_i$ in (\ref{Ki}),
we conclude that $K_i$ has the expansion formula (\ref{Ki-expand}).
Notice Lemma \ref{lemma-Ki} and Lemma \ref{lemma-TKi},
we can prove (\ref{zeta}) by the mathematical induction.

\begin{lemma}
\label{lemma-beta}
Define the combination coefficients 
$\beta_{k,i}$, $0\leq i \leq k+1$, $k=0,1,...$, as follows
	$$ \beta_{0,0}=2, \qquad \beta_{0,1}=1, \qquad
	\beta_{1,0}=-2, \qquad \beta_{1,1}=2, \qquad \beta_{1,2}=1,$$
\begin{equation}
\label{beta}
	\beta_{k+1,i}=\beta_{k,i-1}, \quad 1\leq i \leq k+2, \qquad
	\beta_{k+1,0}=\sum_{i=0}^{k+1}\beta_{k,i} \delta_{i+1}, 
\end{equation}
then we have 
	$$ \sum_{i=0}^{k+1} \beta_{k,i}\alpha_i=0,
	\qquad \sum_{i=0}^{k+1} \beta_{k,i}\gamma_i=0, 
	\qquad k=1,2,.... $$
\end{lemma}

{\parindent=0pt\it Proof.} 
It is easy to check the case when $k=1$.
If the lemma is true for $k$, then
	$$ \sum_{i=0}^{k+1} \beta_{k,i} K_i	
	= O(h) \left(\begin{array}{c} 1 \\ 1 \end{array} \right), $$
so according to Lemma \ref{lemma-TKi}, we have
	$$ \sum_{i=0}^{k+2} \beta_{k+1,i} K_i
	= \widetilde JG \sum_{i=0}^{k+1} \beta_{k,i} K_i 
	= O(h) \left(\begin{array}{c} 1 \\ 1 \end{array} \right), $$
which completes the proof.	

\begin{lemma}
Let $\beta_{k,i}$ be defined by (\ref{beta}). Then we have
\label{lemma-beta-zeta}
\begin{equation}
\label{beta-zeta}
	\sum_{i=0}^{k+1}\beta_{k,i}(\zeta_{i,2}-\zeta_{i,1})
	=(-4)^{k}, \qquad k=1,2,....
\end{equation}
\end{lemma}

{\parindent=0pt\it Proof.} It is easy to check the case when $k=1$.
If the lemma is true for $k$, 
then we have (according to Lemma \ref{lemma-Ki} 
and Lemma \ref{lemma-Ki-expand})
\begin{eqnarray*}
	 \sum_{i=0}^{k+2}\beta_{k+1,i}(\zeta_{i,2}-\zeta_{i,1}) 
	& = & \frac{1}{2}\sum_{i=0}^{k+1}\beta_{k,i}\delta_{i+1}	
	+\sum_{i=1}^{k+2}\beta_{k,i-1}(\zeta_{i,2}-\zeta_{i,1}) \\
	& = & \frac{1}{2}\sum_{i=0}^{k+1}\beta_{k,i}\delta_{i+1}	
	+\sum_{i=0}^{k+1}\beta_{k,i}(-4\zeta_{i,2}+4\zeta_{i,1}
	-\frac{1}{2}\delta_{i+1}+2\gamma_i) \\
	& = & -4\sum_{i=0}^{k+1}\beta_{k,i}(\zeta_{i,2}-\zeta_{i,1})
	+ 2\sum_{i=0}^{k+1}\beta_{k,i}\gamma_i, 
\end{eqnarray*}	
Note Lemma \ref{lemma-beta}, and the proof is completed.

\begin{prop}
\label{prop-Phi}
Given an integer $m>0$,  
let $\beta_{k,i}$ be defined by (\ref{beta}), and set
	$$ \Phi_{2k-1}= (-1)^k 2^{-2k-1}
	\left[-\frac{1}{2}\partial P_k+2\sum\limits_{i=0}^{k+1} 
	\beta_{k,i}(\Psi_{i,1,2k-1}-\Psi_{i,2,2k-1})\right], $$
\begin{eqnarray}
\label{Phi}
	\Phi_{2k}& = & (-1)^k 2^{-2k-1}
	\left[\frac{1}{2}P_{k+1}-(\frac{1}{2}\partial^2
	+\frac{3}{2}q)\frac{1}{2}P_k
	 -\partial\sum\limits_{i=0}^{k+1} 
	\beta_{k,i}(\zeta_{i,2}\Phi_{2k-1}+\Psi_{i,2,2k-1})\right. 
	\nonumber\\
	& & \left. +2\sum\limits_{i=0}^{k+1} 
	\beta_{k,i}(\Psi_{i,1,2k}-\Psi_{i,2,2k})\right], 
\end{eqnarray}
for $k=1,2,...,m-1.$ 
Then under the relation (\ref{w-v-q}) with $L=2m-2$,  $f(x,t)=\frac 12 q(x,t)$ and (\ref{Phi}) 
we have
\begin{equation}
\label{VFP1}
        \widetilde P_m \equiv  \sum_{i=0}^{m+1} \beta_{m,i} K_i =
        \frac{1}{2}  P_m h^{2m}
        \left( \begin{array}{c} 1 \\ 1 \end{array} \right)
        + O(h^{2m+1}), 
\end{equation}
and
\begin{equation}
\label{EQNP1}
        {\left( \begin{array}{c} w \\ v \end{array} \right)}_{t_m}
        +\frac{1}{h^{2m-1}} J \widetilde P_m =
        \frac{1}{2}(   q_{t_m} - B_0   P_{m} ) h^2
        \left( \begin{array}{c} 1 \\ 1 \end{array} \right)
        + O(h^3).
\end{equation}
\end{prop}

{\parindent=0pt\it Proof. } It is easy to check the case when $m=1$,
If the equation (\ref{VFP1}) is valid for $m$, then we have
(according to Lemma \ref{lemma-Ki-expand})
\begin{eqnarray}
	T \widetilde P_m & = & 
	\widetilde JG \sum_{i=0}^{m+1} \beta_{m,i} K_i \\
	& = & \widetilde JG 
	\left[\frac{1}{2}  P_m h^{2m}
        \left( \begin{array}{c} 1 \\ 1 \end{array} \right) 
	+h^{2m+1}\sum_{i=0}^{m+1} \beta_{m,i}
	\left( \begin{array}{c} 
	\zeta_{i,1}\Phi_{2m-1}+\Psi_{i,1,2m-1} \\
	\zeta_{i,2}\Phi_{2m-1}+\Psi_{i,2,2m-1}  \end{array} \right)\right. \nonumber\\
	& & \left. +h^{2m+2}\sum_{i=0}^{m+1} \beta_{m,i}
	\left( \begin{array}{c}  
	\zeta_{i,1}\Phi_{2m}+\Psi_{i,1,2m} \\
	\zeta_{i,2}\Phi_{2m}+\Psi_{i,2,2m}  \end{array} \right) 
        + O(h^{2m+3})\right],	
\end{eqnarray}
note the definition of $\Phi_{2m-1}$ and $\Phi_{2m}$ in (\ref{Phi}), 
we obtain (due to (\ref{beta-zeta}))
\begin{equation}
\label{2k+1}
	-2 \sum_{i=0}^{m+1} \beta_{m,i}
	(\zeta_{i,1}\Phi_{2m-1}+\Psi_{i,1,2m-1}) 
	+ 2 \sum_{i=0}^{m+1} \beta_{m,i}
	(\zeta_{i,2}\Phi_{2m-1}+\Psi_{i,2,2m-1}) + \frac{1}{2}\partial P_m = 0,
\end{equation}
and 
	$$ (\frac{1}{2}\partial^2+\frac{3}{2}q)\frac{1}{2}P_m
	+\partial \sum_{i=0}^{m+1} \beta_{m,i}(\zeta_{i,2}\Phi_{2m-1}+\Psi_{i,2,2m-1}) $$
\begin{equation}
\label{2k+2a}
	-2\sum_{i=0}^{m+1} \beta_{m,i}(\zeta_{i,1}\Phi_{2m}+\Psi_{i,1,2m})
	+ 2 \sum_{i=0}^{m+1} \beta_{m,i}(\zeta_{i,2}\Phi_{2m}+\Psi_{i,2,2m}) 
	= \frac{1}{2}P_{m+1}.
\end{equation}
Combining the above two equations (\ref{2k+1}) and (\ref{2k+2a}), 
and noting the equation (\ref{KdVBIHS}), we have
	$$ (\frac{1}{2}\partial^2-\frac{1}{2}\partial^{-1}q_x
	+\frac{1}{2}\partial^{-1}q\partial)\frac{1}{2}P_m
	-\partial \sum_{i=0}^{m+1} \beta_{m,i}(\zeta_{i,1}\Phi_{2m-1}+\Psi_{i,1,2m-1}) $$
\begin{equation}
\label{2k+2b}
	+2\sum_{i=0}^{m+1} \beta_{m,i}(\zeta_{i,1}\Phi_{2m}+\Psi_{i,1,2m})
	- 2 \sum_{i=0}^{m+1} \beta_{m,i}(\zeta_{i,2}\Phi_{2m}+\Psi_{i,2,2m}) 
	= \frac{1}{2}P_{m+1}.
\end{equation}
So we get 
\begin{equation}
	T \widetilde P_m = \frac{1}{2}  P_{m+1} h^{2m+2}
        \left( \begin{array}{c} 1 \\ 1 \end{array} \right)
        + O(h^{2m+3}).
\end{equation}
On the other hand (according to Lemma \ref{lemma-TKi}), 
\begin{equation}
	T \widetilde P_m  =  \widetilde JG \sum_{i=0}^{m+1} \beta_{m,i} K_i 
	= \sum_{i=0}^{m+1}  \beta_{m,i}( K_{i+1} + \delta_{i+1} K_0) 
	= \widetilde P_{m+1}.
\end{equation}
The equation (\ref{EQNP1}) is the corollary of the equation (\ref{VFP1})
and Proposition \ref{prop-Poisson}.
The proof is finished.

We give an example here.
For $m=3$,  using Proposition \ref{prop-Phi}, we can get 
\begin{equation}
	\Phi_1 = \frac{1}{8} q_x, \qquad \Phi_2 = -\frac{1}{32}q^2,
	\qquad \Phi_3=-\frac{1}{384}q_{xxx}, \qquad
	\Phi_4=\frac{1}{254}(q^3+qq_{xx}+q_x^2),
\end{equation}
then under the relation (\ref{w-v-q}) with $L=4$,  $f(x,t)=\frac 12 q(x,t)$ and the above 
$\Phi_i$'s
we have
	$$ -10 K_0 + 4 K_1 -2 K_2 + 2 K_3 +  K_4 =  \frac{1}{2}  P_3 h^6 
        \left( \begin{array}{c} 1 \\ 1 \end{array} \right) + O(h^7). $$
In the previous work in \cite{ZLC}, 
we must combine $K_0$, $K_1$, ..., $K_6$ for giving $P_3$ under the relation (\ref{w-v-q}) with 
$L=0$.
In general, $K_0$, $K_1$, ..., $K_{2m}$ are needed to be combined for giving $P_m$
under the relation (\ref{w-v-q}) with $L=0$ \cite{ZLC}.
Proposition \ref{prop-Phi} shows us that almost only half of them, 
i.e., $K_0$, $K_1$, ..., $K_{m+1}$, are needed to give $P_m$
by introducing $\Phi_i(f)$ (\ref{Phi}).
Furthermore, according to the recursion formula for $\Phi_i(f)$ (\ref{Phi})
it is easy to see that all the $\Phi_i(f)$'s, introduced by (\ref{Phi}), 
are differential polynomials of $f$,
and our process for finding $\Phi_i(f)$ can be continued indefinitly.

In what follows, we will derive the continuous limit relations
between the Hamiltonians, the Lax pairs of the Toda hierarchy
and those of the KdV hierarhcy,  respectively.

\begin{lemma}
\label{lemma-tilde-w-q}
If there is a relation between 
$\widetilde w(n)$, $n\in\mathbb Z$, and $q(x)$, $x\in\mathbb R$
\begin{equation}
\label{tilde-w-q}
	\widetilde w(n)=q^{(s_1)}(x)q^{(s_2)}(x)\cdots q^{(s_m)}(x) h^l,
\end{equation}
where $h$ is the step of lattice, $x=nh$,
$s_i$, $1\leq i\leq m$ and $l$ are nonnegtive integers,
and denote $\widetilde S$ as the operator 
which stands for submitting the relation (\ref{tilde-w-q}) 
into a polynomial of $\widetilde w$, $\widetilde w^{(-1)}$, $\widetilde w^{(1)}$, ...,
and then expanding in Taylor series, 
then we have the formula
\begin{equation}
	\frac{\delta }{\delta q} \circ \widetilde S
	= h^l \widetilde Z \circ \widetilde S\circ \frac{\delta}{\delta \widetilde w}, 
\end{equation}
where $\widetilde Z$ stands for a differential operator.
\end{lemma}

The proof for Lemma \ref{lemma-tilde-w-q} is given in Appendix A.

\begin{prop}
\label{prop-Hk}
Given an integer $m>0$,  
set
\begin{equation}
	\widetilde H_m \equiv \sum_{i=0}^{m+1} \beta_{m,i} H_i 
	- \sum_{i=1}^{m+1} \beta_{m,i} \frac{\alpha_{i+1}}{i},
\end{equation}
under the relation (\ref{w-v-q}) with $L=2m-2$, $f(x,t)=\frac 12 q(x,t)$ and (\ref{Phi}),
we have
\begin{equation}
	\int S(\widetilde H_m) dx
	= \frac{1}{2}h^{2m+2}\int \overline H_m dx +O(h^{2m+3}), 
\end{equation}
where $S$ is an operator 
which stands for submitting the relation (\ref{w-v-q})  
with $L=2m-2$,  $f(x,t)=\frac 12 q(x,t)$ and (\ref{Phi}) 
into a polynomial of $w$, $v(n)$, $w^{(-1)}$, $v^{(-1)}$,  $w^{(1)}$, $v^{(1)}$, ...,
and then expanding in Taylor series. 
\end{prop}

{\parindent=0pt\it Proof.} 
According to Lemma \ref{lemma-tilde-w-q},
under the relation (\ref{w-v-q})  
with $L=2m-2$,  $f(x,t)=\frac 12 q(x,t)$ and (\ref{Phi}),
(since $\Phi_i$'s are differential polynomials of $q$),
we have
\begin{eqnarray*}
	\frac{\delta}{\delta q} \circ S 
	& = & \sum_{j=0}^\infty (-\partial)^j 
	\frac{\partial}{\partial q^{(j)}} 	\circ S \\
	& = & \sum_{j=0}^\infty (-\partial)^j \sum_{k\in\mathbb Z} 
	\left[(\frac{\partial S(w^{(k)})}{q^{(j)}}) S\circ 
	\frac{\partial}{\partial w^{(k)}} 
	 + (\frac{\partial S(v^{(k)})}{q^{(j)}}) S\circ 
	\frac{\partial}{\partial v^{(k)}}\right] \\
	& = & \frac{1}{2} h^2 \sum_{j=0}^\infty (-\partial)^j \sum_{k\in\mathbb Z} 
	\frac{(kh)^{j}}{j!} S \circ (\frac{\partial}{\partial w^{(k)}}+
	\frac{\partial}{\partial v^{(k)}}) 
	 + h^3 Z\circ S \circ (\frac{\delta}{\delta w}-\frac{\delta}{\delta v}) \\
	& = & \frac{1}{2} h^2 S \circ (\frac{\delta}{\delta w}+\frac{\delta}{\delta v}) 
	+ h^3 Z\circ S \circ (\frac{\delta}{\delta w}-\frac{\delta}{\delta v}),
\end{eqnarray*} 
where $Z$ stands for a differential operator, 
and we do not care about its concrete form.
Note Lemma \ref{lemma-Ki} and 
the definition of $H_i$ in (\ref{Ki}), 
we can have the expansion
	$$ S(\widetilde H_m) = \sum_{i=2}^\infty \widetilde H_{m,i} h^i, $$
where $\left. \widetilde H_{m,i} \right|_{q=0} =0$, 
and according to Proposition \ref{prop-Phi},
we have
\begin{eqnarray*}
	\frac{\delta}{\delta q} \circ S  (\widetilde H_m) 
	& = &\sum_{i=2}^\infty h^i\frac{\delta \widetilde H_{m,i}}{\delta q} \\
	&  = & \left[ \frac{1}{2} h^2 S \circ (\frac{\delta}{\delta w}+
	\frac{\delta}{\delta v}) 
	+ h^3 Z \circ S \circ (\frac{\delta}{\delta w}-\frac{\delta}{\delta v}) \right]
	\sum_{i=0}^{m+1} \beta_{m,i} H_m \\
	& = & \frac{1}{2} h^{2m+2} \frac{\delta \overline H_m}{\delta q} 
	+ O(h^{2m+3}). 
\end{eqnarray*}
Then one can get \cite{TUTD}
	$$ \widetilde H_{m,i} \in {\rm Const.} + \mbox{Image}(\partial), 
	\qquad 2\leq i \leq 2m+1. $$
As we mentioned above, 
there is no constant item in each $\widetilde H_{m,i}$, $i\geq 2$,
(i.e., $\left. \widetilde H_{m,i} \right|_{q=0} =0$), so
	$$ \int \widetilde H_{m,i} dx = 0, \qquad  2\leq i \leq 2m+1. $$
Just using the same deduction, we conclude
	$$ \int \widetilde H_{m,2m+2} dx = \frac{1}{2} \int \overline H_m dx, $$
which completes the proof.

\begin{lemma}
\label{lemma:A_m-expand}
Under the relation (\ref{w-v-q}) with $f(x,t)=\frac 12 q(x,t)$,
we have
\begin{equation}
\label{A_m-expand}
	A_k = \alpha_k-\gamma_k+
	\sum_{i=2}^\infty A_{k,i} h^i ,
	\qquad k=0,1,...,	
\end{equation}	
where
\begin{equation}	
	A_{k,2i}|_{q=0} =0,\quad
	A_{k,2i+1}|_{q=0}
	=\xi_{k,2i+1} \partial^{2i+1},
	\qquad i=1,2,...,	
\end{equation}
$\xi_{k,2i+1}$is a constant,
and $\alpha_k$ and $\gamma_k$
are given in Lemma \ref{lemma-Ki}.
\end{lemma}

{\parindent=0pt\it Proof.} For $k=0$ and $k=1$, we have
	$$A_0|_{q=0}=-\frac{1}{2},
	\qquad A_1|_{q=0}
	=1+\sum_{j=0}^\infty \frac{1}{(2j+1)!}
	h^{2j+1}(-\partial)^{2j+1}.$$
If the lemma is valid for $k-1$,
note $\alpha_k=-2\alpha_{k-1}+2\gamma_{k-1}$
(see Lemma \ref{lemma-Ki}), we have
\begin{eqnarray*}
	A_k|_{q=0} 
	&=& \left. A_{k-1} (E + w + v E^{(-1)}) 
	- vb_k^{(1)}E^{(-1)}
	-a_k \right|_{q=0} \\
	&=& \left[\alpha_{k-1}-\gamma_{k-1}+
	\sum_{i=0}^\infty \xi_{k-1,2i+1} 
	h^{2i+1} \partial^{2i+1}\right] 
	\sum_{j=1}^\infty 
	\frac{2}{(2j)!}h^{2j}\partial^{2j} \\
	& & +\alpha_k\sum_{j=0}^\infty
	\frac{1}{j!}h^j(-\partial)^j -\gamma_k \\
	&\equiv& \alpha_k-\gamma_k+
	\sum_{i=0}^\infty \xi_{k,2i+1} 
	h^{2i+1} \partial^{2i+1}. 
\end{eqnarray*}

\begin{lemma}
\label{lemma-tilde-A_m}
Define
\begin{equation}
	\widetilde A_k \equiv
	\sum_{i=1}^{k+1} \beta_{k,i} A_{i-1}
	\qquad k=1,2,....
\end{equation}
Then under the relation (\ref{w-v-q}) with $f(x,t)=\frac 12 q(x,t)$,
we have
\begin{equation}
	\widetilde A_k =\sum_{i=2}^\infty 
	\widetilde A_{k,i} h^{i},
\end{equation}
where
\begin{equation}
	\widetilde A_{k,2i}|_{q=0}=0,\quad
	\widetilde A_{k,2i+1}|_{q=0} =
	\widetilde \xi_{k,2i+1} \partial^{2i+1},
	\qquad i=1,2,...,	
\end{equation}
$\widetilde \xi_{k,2i+1}$is a constant.
\end{lemma}

{\parindent=0pt\it Proof.}
According to Lemma \ref{lemma:A_m-expand},
we only need to prove
\begin{equation}
	\sum_{i=1}^{k+1} \beta_{k,i} 
	(\alpha_{i-1}-\gamma_{i-1})=0.
\end{equation}
It is easy to check the cases: $k=1$ and $k=2$,
and for $k\geq 3$, note Lemma \ref{lemma-beta}, we have
	$$ \sum_{i=1}^{k+1}\beta_{k,i}(\alpha_{i-1}-\gamma_{i-1})
	=\sum_{i=1}^{k+1}\beta_{k-1,i-1} (\alpha_{i-1}-\gamma_{i-1})
	=\sum_{i=0}^k \beta_{k-1,i}(\alpha_{i}-\gamma_{i})=0, $$
which completes the proof.

\begin{prop}
\label{prop-Ak}
Given an integer $m>0$,
under the relation (\ref{w-v-q}) with $L=2m-2$, $f(x,t)=\frac 12 q(x,t)$ and (\ref{Phi}),
we have
\begin{equation}
	\widetilde A_m \equiv \sum_{i=1}^{m+1} \beta_{m,i} A_{i-1}
	= - \overline A_m h^{2m-1}+O(h^{2m}). 
\end{equation}
\end{prop}

{\parindent=0pt\it Proof.} 
It is valid for $m=1,2$.
According to Proposition \ref{prop-Phi},  
we have
\begin{eqnarray}
\label{widetilde-A_k-L}
	[ \widetilde A_m, L] 
	&=& \sum_{i=1}^{m+1} \beta_{m,i}\frac{d w}{d t_{i-1}} 
	+\sum_{i=1}^{m+1} \beta_{m,i}\frac{d v}{d t_{i-1}} E^{(-1)} 
	\nonumber \\
	&=& J_{12}\sum_{i=1}^{m+1} \beta_{m,i} K_{i,2}+
	J_{21}\sum_{i=1}^{m+1}\beta_{m,i} K_{i,1} E^{(-1)} 
	\nonumber \\
	&=& - B_0 P_m h^{2m+1}+O(h^{2m+2}) \nonumber \\
	&=& -[ \overline A_m, \overline L]h^{2m+1}+O(h^{2m+2}).
\end{eqnarray}
Under the relation (\ref{w-v-q}) with $L=2m-2$, $f(x,t)=\frac 12 q(x,t)$ and (\ref{Phi}),
Proposition \ref{prop-spectral} and Lemma \ref{lemma-tilde-A_m} together imply 
\begin{equation}
	L = \overline L h^2+\sum_{i=3}^\infty L_i h^i,
	\quad\qquad
	\widetilde A_m = \sum_{i=2}^\infty 
	\widetilde A_{m,i} h^i,
\end{equation}
where $L_i$ and $\widetilde A_{m,i}$ are differential operators.
Comparing the terms of $h^4$ in (\ref{widetilde-A_k-L}),
we know
\begin{equation}
	[\widetilde A_{m,2}, \overline L]=0,
\end{equation}
According to \cite{Drinfeld-Sokolov-85}, 
$\widetilde A_{m,2}$ can be represented by
\begin{equation}
	\widetilde A_{m,2}=\sum_{j=0}^\infty
	\eta_{m,2,j} (\overline L)^j,
\end{equation}
where $\eta_{m,2,j}$ are constants.
Note Lemma \ref{lemma-tilde-A_m}, we have
\begin{equation}
	\widetilde A_{m,2}|_{q=0}=0
	=\sum_{j=0}^\infty
	\eta_{m,2,j} (\partial^2)^j.
\end{equation}
Then one can get $\eta_{m,2,j}=0$ for all $j$, and 
\begin{equation}
\label{A_m_2}
	\widetilde A_{m,2} =0.
\end{equation}
Comparing the terms of $h^5$ in (\ref{widetilde-A_k-L}),
we know
\begin{equation}
	[\widetilde A_{m,3}, \overline L]=0,
\end{equation}
then $\widetilde A_{m,3}$ can be represented by
\cite{Drinfeld-Sokolov-85}
\begin{equation}
	\widetilde A_{m,3}=\sum_{j=0}^\infty
	\eta_{m,3,j} (\overline L)^j,
\end{equation}
where $\eta_{m,3,j}$ are constants.
Note Lemma \ref{lemma-tilde-A_m},
and we have
\begin{equation}
	\widetilde A_{m,3}|_{q=0}
	=\widetilde \xi_{m,3}\partial^3
	=\sum_{j=0}^\infty
	\eta_{m,3,j} (\partial^2)^j.
\end{equation}
Then one can get $\eta_{m,3,j}=0$ for all $j$, and
\begin{equation}
	\widetilde A_{m,3} =0.
\end{equation}
In the same way, we conclude 
\begin{equation}
	\widetilde A_{m,i} =0, \qquad
	i=2,...,2m-2.
\end{equation}
Comparing the terms of $h^{2m+1}$ in (\ref{widetilde-A_k-L}),
we know
\begin{equation}
	[\widetilde A_{m,2m-1}, \overline L]
	=-[\overline A_m, \overline L],
\end{equation}
then $\widetilde A_{m,2m-1}+\overline A_m$ can be represented by
\cite{Drinfeld-Sokolov-85}
\begin{equation}
	\widetilde A_{m,2m-1}+\overline A_m
	=\sum_{j=0}^\infty
	\eta_{m,2m-1,j} (\overline L)^j,
\end{equation}
where $\eta_{m,2m-1,j}$ are constants.
Note Lemma \ref{lemma-tilde-A_m} and (\ref{KdV-Lax-t}), we have
\begin{equation}
	\left. \left( \widetilde A_{m,2m-1}+\overline A_m
	\right) \right|_{q=0}
	=\widetilde \xi_{m,2m-1} \partial^{2m-1}
	+\partial^{2m-1}
	=\sum_{j=0}^\infty
	\eta_{m,2m-1,j} (\partial^2)^j.
\end{equation}	
Then we get $\eta_{m,2m-1,j}=0$ for all $j$ and 
\begin{equation}
	\widetilde A_m \equiv
	\sum_{i=1}^{2m} \beta_{m,i} A_{i-1}
	= - \overline A_m h^{2m-1} +O(h^{2m}).
\end{equation}
Thus the proof is completed.

\section{Conclusions and remarks}

In this paper, by introducing the higher order terms in the potential expansion,
we have proved that there is the continuous limit relation 
between the Toda hierarchy and the KdV hierarchy.
Compared with the \cite{ZLC},
the fewer members of the Toda hierarchy are needed 
to recover the KdV hierarchy by the recombination method.
For example, Proposition \ref{prop-Phi} shows that
under the potential expansion (\ref{w-v-q}) 
with $f(x,t)=\frac 12 q(x,t)$ and (\ref{Phi}),
we can combine $K_0$, $K_1$, $...$, $K_{m+1}$, 
to get $P_m$ in continuous limit.
However, under the lower finite potential expansion,
for example (\ref{w-v-q}) with $f(x,t)=\frac 12 q(x,t)$ and $L=0$, 
we need $K_0$, $K_1$, $...$, $K_m$, $...$, $K_{2m}$, 
to recover $P_m$ through the continuous limit process \cite{ZLC}.

Compared with the \cite{Gieseker},
a new method  for introducing $\Phi_i(f)$ in the potential expansion (\ref{w-v-q}) 
was presented in this paper. 
Moreover, from the recursion formula for $\Phi_i(f)$ (\ref{Phi}),
it is easy to see that the $\Phi_i(f)$'s, introduced in our construction,
are all differential polynomials of $f$,
and our process for determining $\Phi_i(f)$ can be continued indefinitly.
However, this can not be obtained in \cite{Gieseker}, 
since the $\Phi_i(f)$'s are obtained by integration there.

It was also shown that the Lax pairs, the Poisson tensors, 
and the Hamiltonians of the Toda hierarchy
tend towards the corresponding ones of the KdV hierarchy in continuous limit

\section*{Acknowledgements}

This work was in part supported by 
the City University of Hong Kong (Project Nos.: 7001178, 7001041),
the RGC of Hong Kong (Project No.: 9040466),
and Chinese Basic Research Project ``Nonlinear Science".
One of the authors (R.L. Lin) acknowledges warm hospitality 
at the City University of Hong Kong.

\section*{Appendix A. Proof of Lemma \ref{lemma-tilde-w-q}}

Denote $\widetilde w_i=q^{(s_1)}\cdots q^{(s_{i-1})}q^{(s_{i+1})}\cdots q^{(s_m)},$ 
for $i=1,...,m$, then we have
\begin{eqnarray*}
	\frac{\delta }{\delta q} \circ \widetilde S
	& = & \sum_{j=0}^\infty 
	(-\partial)^j \frac{\partial}{\partial q^{(j)}} \circ \widetilde S \\
	& = & \sum_{j=0}^\infty (-\partial)^j 
	\sum_{k\in{\mathbb Z}} 
	\left( \frac{\partial \widetilde S(\widetilde w^{(k)})}{\partial q^{(j)}} \right)
	\widetilde S\circ\frac{\partial}{\partial \widetilde w^{(k)}} \\
	& = & h^l \sum_{i=1}^m \sum_{j=s_i}^\infty (-\partial)^j 
	\sum_{k\in{\mathbb Z}} \frac{(kh)^{j-s_i}}{(j-s_i)!}
	\left( e^{kh\partial} \widetilde S(\widetilde w_i) \right) 
	\widetilde S\circ\frac{\partial}{\partial \widetilde w^{(k)}} 	\\
	& = & h^l \sum_{i=1}^m \sum_{j=0}^\infty (-\partial)^{j+s_i} 
	\sum_{k\in{\mathbb Z}} \frac{(kh)^{j}}{j!}
	\left( e^{kh\partial} 	\widetilde S(\widetilde w_i)  \right) 
	\widetilde S\circ\frac{\partial}{\partial \widetilde w^{(k)}} \\	
	& = & h^l \sum_{i=1}^m (-\partial)^{s_i} 
	\sum_{j=0}^\infty \sum_{k\in{\mathbb Z}} 
	\sum_{p=0}^j \frac{(-kh)^{j}}{p! (j-p)!} 
	\left( \partial^p e^{kh\partial} \widetilde S(\widetilde w_i)  \right) 
	\partial^{j-p}\circ \widetilde S\circ\frac{\partial}{\partial \widetilde w^{(k)}} \\
	& = & h^l \sum_{i=1}^m (-\partial)^{s_i} \sum_{p=0}^\infty
	 \sum_{k\in{\mathbb Z}} \sum_{j=p}^\infty
	 \frac{(-kh)^{j}}{p! (j-p)!} 
	\left( \partial^p e^{kh\partial} \widetilde S(\widetilde w_i) \right) 
	\partial^{j-p}\circ \widetilde S\circ\frac{\partial}{\partial \widetilde w^{(k)}} 	
\\
	& = & h^l \sum_{i=1}^m (-\partial)^{s_i} \sum_{p=0}^\infty
	\left( \partial^p e^{kh\partial} \widetilde S(\widetilde w_i) \right)
	 \sum_{k\in{\mathbb Z}} \sum_{j=0}^\infty
	 \frac{(-kh)^{j+p}}{p! j!} 	 
	\partial^{j}\circ \widetilde S\circ\frac{\partial}{\partial \widetilde w^{(k)}} 	
\\
	& = & h^l \sum_{i=1}^m (-\partial)^{s_i} \sum_{p=0}^\infty
	\frac{(-kh)^{p}}{p! }
	\left( \partial^p e^{kh\partial} \widetilde S(\widetilde w_i) \right)
	 \sum_{k\in{\mathbb Z}} \sum_{j=0}^\infty
	 \frac{(-kh)^{j}}{j!} 	 
	\partial^{j}\circ \widetilde S\circ\frac{\partial}{\partial \widetilde w^{(k)}} \\	
	& = & h^l \sum_{i=1}^m (-\partial)^{s_i} \sum_{p=0}^\infty
	\frac{(-kh)^{p}}{p! }
	\left( \partial^p e^{kh\partial} \widetilde S(\widetilde w_i)  \right)
	\widetilde S\circ \sum_{k\in{\mathbb Z}} E^{(-k)}
	\circ\frac{\partial}{\partial \widetilde w^{(k)}} \\
	& = & h^l \sum_{i=1}^m (-\partial)^{s_i} \sum_{p=0}^\infty
	\frac{(-kh)^{p}}{p! }
	\left( \partial^p e^{kh\partial} \widetilde S(\widetilde w_i) \right)
	\widetilde S \circ\frac{\delta}{\delta \widetilde w^{(k)}} \\
	& \equiv & h^l \widetilde Z \circ \widetilde S\circ \frac{\delta}{\delta \widetilde w}.
\end{eqnarray*}
The proof for Lemma \ref{lemma-tilde-w-q} is finished.


\begin{thebibliography}{s99}

\bibitem{Toda-73}
Toda M and Wadati M 
1973 A soliton and two solitons in an exponential lattice and related equations
{\it J. Phys. Soc. Japan} {\bf 34} 18-25

\bibitem{Case-74}
Case K M and Kac M. 
1974 A discrete version of the inverse scattering problem
{\it J. Math. Phys.} {\bf 14} 594-603

\bibitem{Kupershmidt-85}
Kupershmit B A 
1985 Discrete Lax equations and differential-difference calculus
{\it Ast\'erisque} {vol 123} (Paris: Soc. Math. France)

\bibitem{Ablowitz-81}
Ablowitz M J and Segur H
1981  Solitons and the inverse scattering transform
(Philadelphia: SIAM)

\bibitem{Zeng-95a}
Zeng Y B and Rauch-Wojciechowski S
1995 Restricted flows of the Ablowitz-Ladik hierarchy and their continuous limits
{\it J. Phys. A:Math. Gen.} {\bf 28} 113-134

\bibitem{Zeng-95b}
Zeng Y B and Rauch-Wojciechowski S
1995 Continuous limit for the Kac-van Moerbeke hierarchy and
for their restricted flows
{\it J. Phys. A: Math. Gen.} {\bf 28} 3825-3840

\bibitem{Morosi-96}
Morosi C and Pizzocchero  L
1996 On the continuous limit of integrable lattices I. 
The Kac-Moerbeke system and KdV theory
{\it Commun. Math. Phys.} {\bf 180} 505-528

\bibitem{Morosi-98a}
Morosi C and Pizzocchero L
1998 On the continuous limit of integrable lattices II.
Volterra system and $sp(N)$ theories
{\it Rev. Math. Phys.} {\bf 10} 235-270

\bibitem{Morosi-98b}
Morosi C and Pizzocchero L
1998 On the continuous limit of integrable lattices III.
Kupershmidt systems and $sl(N+1)$ KdV theories
{\it J. Phys. A: Math. Gen.} {\bf 31} 2727-2746

\bibitem{Gieseker} Gieseker D
1996 The Toda hierarchy and the KdV hierarchy  
{\it Commun. Math. Phys.} {\bf 181} 587-603

\bibitem{ZLC} Zeng Y B,  Lin R L and Cao  X
1999 The relation between the Toda hierarchy and the KdV hierarchy  
{\it Phys. Lett.} {\bf 251A} 177-183

\bibitem{TUTD} Tu Gui-zhang 
1990 A trace identity and its applications to the theory of discrete integrable systems  
{\it J. Phys. A: Math. Gen.} {\bf 23} 3903-3922

\bibitem{Newell} 
Newell A C 
1985  Soliton in mathematics and physics
(Philadelphia: SIAM)

\bibitem{Ma-JMP} Ma W X and Fuchssteiner  B
1999 Algebraic structure of discrete zero curvature equations 
and master symmetries of discrete evolution equations 
{\it J. Math. Phys.} {\bf 40} 2400-2418

\bibitem{Ma-99} Fuchssteiner B and Ma W X   
1999
``An approach to master symmetries of lattice equations",
{\it Symmetries and Integrability of Difference Equations},
247-260, edited by Clarkson P A and Nijhoff F W 
(Cambridge: London Math. Soc.) pp 247-260

\bibitem{Drinfeld-Sokolov-85} 
Drinfeld V G and Sokolov V V  
1985 Lie algebras and equations of Korteweg-de Vries type
{\it J. Sov. Math.} {\bf 30} 1975-2036

\end {thebibliography}
\end{document}